\documentstyle[12pt]{article}
\newcommand{\Section}[1]{\section{#1}\setcounter{equation}{0}}

\newcommand{\be}{\begin{equation}}
\newcommand{\bea}{\begin{eqnarray}}
\newcommand{\eea}{\end{eqnarray}}
\newcommand{\ba}{\begin{array}}
\newcommand{\ea}{\end{array}}
\newcommand{\ee}{\end{equation}}

\newcommand{\qq}{\mbox{\boldmath$q$}}
\newcommand{\mm}{\mbox{\boldmath$g$}}
\expandafter\ifx\csname mathbbm\endcsname\relax

\newcommand{\bone}{\mbox{1\hspace{-0.65em}1}}
\else

\newcommand{\bone}{{\mathbbm{1}}}
\fi
\def\l{\label}
\def\o{\over}
\def\p{\partial}
\def\a{\alpha}
\def\1{a _1}
\def\2{a _2}
\def\3{a _3}
\def\4{a _4}
\def\5{a _5}
\def\6{a _6}
\def\7{a _7}
\def\8{a _8}
\def\nn{\nonumber}

\def\l{\label}
\begin{document}
\begin{titlepage}
\hfill
\vbox{
    \halign{#\hfil         \cr
           IPM-96-144   \cr
           May 1996   \cr
           } 
      }  
\vspace*{3mm}
\begin{center}
{\LARGE The Moduli Space and Monodromies of the $N=2$ Supersymmetric
Yang-Mills Theory with
any Lie Gauge Groups\\}
\vspace*{20mm}
{\large M. R. Abolhasani$^{+,}$\footnote{e-mail:hasani@physic.sharif.ac.ir}
,\,
\,
M. Alishahiha$^{*,+,}$\footnote{e-mail:alishah@physics.ipm.ac.ir},\,\,
A. M. Ghezelbash$^{*,\dagger ,}$\footnote{e-mail:amasoud@physics.ipm.ac.ir}}
\\
{\it $^{*}$Institute for Studies in Theoretical Physics and Mathematics, \\
 P.O.Box 19395-1795, Tehran, Iran } \\
{\it $^{+}$Department of Physics, Sharif University of Technology, \\
\it  P.O.Box 11365-9161, Tehran, Iran }\\
{\it $^{\dagger}$Department of Physics, Alzahra University, \\
\it  Tehran 19834, Iran }\\
\vspace*{25mm}
\end{center}
\begin{abstract}
We propose a unified scheme for finding the hyperelliptic curve of $N=2$
SUSY YM theory with any Lie gauge groups.
Our general scheme gives the well known results for classical gauge groups
and exceptional $G_2$ group.
In particular, we present the curve for the exceptional gauge groups $F_4,
E_{6,7,8}$ and check consistency condition for them.
The exact monodromies and the dyon spectrum of these theories are determined.
We note that for any Lie gauge groups, the exact monodromies could be
obtained only from the Cartan matrix.
\end{abstract}
\end{titlepage}
\newpage
\Section{Introduction}
During the two years ago, exact results on the four dimensional quantum
field theory
have been obtained . Although all of these models are
supersymmetric, the study of them, helps to understanding some results on the
non-perturbative quantum field theory in four dimensions. The key point in
$N=2$ SUSY YM models was the discovery of a hyperelliptic curve which gives
all the informations about low energy effective action. At first,
Seiberg and Witten have noticed to this point for
$N=2$ $SU(2)$ SUSY YM theory
\cite{SW}.

This work have been generalized for $SU(N)$ \cite{LA}, $SO(2n+1)$
\cite{DA},
$SO(2n)$ \cite{BR}, $SP(2n)$ \cite{ARG}
and $G_2$ \cite{ALI} gauge groups. Also $N=2$ supersymmetric gauge
theory with matter multiplet in
the fundamental representation of the gauge group has been considered in
\cite{HAN}.
The low energy effective action for the $N=2$ gauge theory, written in terms
of the $N=1$ fields is:
\be \l{LAG}
{ 1 \over 4 \pi} Im \left( \int d^{4} \theta { \partial {\cal F} (A)
\over \partial A^{i} } \bar{A}^{i} + \int d^{2} \theta {1 \over 2}
{ \partial ^{2} {\cal F} ( A ) \over \partial A^{i} \partial A^{j}}
W^{i}_{\alpha} W^{j}_{\alpha} \right),
\ee
where the $A^{i}$'s are the $N=1$ chiral field multiplets and
$W^{i}_{\alpha}$
are the vector multiplets all in the adjoint representation.
We note first that the whole theory is described by
a single holomorphic function, the prepotential ${\cal F}(A)$  which at the
classical level is ${\cal F}={1 \o 2} \tau A^2$,$\tau =i{{4\pi} \o {g^2}}+
{\theta \o {2\pi}}$. Second, this theory has a classical potential for scalar
fields given by
$V(\phi)=tr {[\phi,\phi^{\dagger}]}^2$.

The v.e.v. of the scalar componet $\phi$ of $A$, determines the moduli space
of the Coulomb phase of the theory, which is in turn parametrized by the
invariants of the gauge group.
\be \l{PHI}
<\phi>=\sum a_{i} H_{i},
\ee
with $H_{i}$'s denoting the Cartan generators.

In a generic point of the moduli space, the gauge group will be broken to
$U(1)^{r}$, where $r$ is the rank of the group
and the W-bosons corresponding to the roots $\vec \alpha$ of the gauge group
become
massive with a mass proportional to $(\vec a\cdot \vec \alpha)^2$.
When the v.e.v of scalar field is orthogonal to a root, the corresponding
W-boson
becomes massless and the symmetry is enhanced, therefore the low energy
effective action
is not valid. In the quantum case, the theory is described by a scale factor
$\Lambda$
, and the prepotential in the limit of weak coupling gets one loop correction
\be \l{F}
{\cal F}={i \o {4\pi}} \sum _{\vec \alpha \in \delta ^+}(\vec \alpha \cdot
\vec
a)^2 \ln {{(\vec \alpha \cdot \vec a)^2} \o {\Lambda ^2}},
\ee
where $\delta ^+$ is the set of positive roots of the gauge group.
If the gauge group is a non-simply laced group, then in the (\ref{F}), $(\vec
\alpha \cdot \vec a)^2$ must be replaced by
${{2(\vec \alpha \cdot \vec a)^2} \o {(\vec \alpha \cdot \vec \alpha)}}$
\cite{ALV}.
One can see that the logarithmic term diverges where
$\vec a\cdot \vec \alpha =0$, which
are exactly the walls of the Weyl chambers in complexified Cartan subalgebra
. The prepotential is not a single valued
function of $\vec a$, therefore encircling the singularities in the moduli
space gives non-trivial monodromy. One can consider the locus of the
singularities to
be the zeros of the Weyl invariant classical discriminant defined by
\be
\Delta _{cl}=\prod _{\vec \alpha \in \delta ^+} (\vec a\cdot \vec \alpha)^2.
\ee
All the semiclassical
monodromies are generated by $r$ simple monodromies corresponding to the $r$
simple roots of the gauge group. The simple monodromy of the simple root
$\alpha _i$ is \cite{DA}
\be      \l{MRT}
M^{(r_i)}=\pmatrix{{(r_i^{-1})}^t&\alpha _i \otimes \alpha _i \cr 0 &r_i},
\ee
where $r_i$ is the Weyl reflection corresponding to the root $\alpha _i$,
and acts on the pair $\pmatrix{\vec a^D \cr \vec a}$ where
$a^D_i={{\p {\cal F}} \o {\p a_i}}$.
Note that the off diagonal terms in (\ref{MRT}) can be changed by a
change of homology
basis ($a^D_i \rightarrow a^D_i+K_{ij}a_j$). One can also
decompose (\ref{MRT}) to a classical part that is generated by Weyl
reflections and a quantum part $M^{(r_i)}=P^{(r_i)}T_i^{-1}$ \cite{LA}.
The important discovery in the $N=2$ gauge theories has been the realization
that the prepotentials can be described with the aid of a family of complex
curves, with the identification of the v.e.v., $a_{i}$ and its dual
$a^{D}_{i}$, with the periods of the curve ,
\be
a_{i} = \oint _{\alpha _{i}} \lambda \;\;\;\; \mbox{and} \;\;\;\;
a^{D}_{i} = \oint _{\beta_{i}} \lambda , \label{integrals}
\ee
where $\alpha_{i}$ and $\beta_{i}$ are the homology cycles of the
corresponding Riemann surface.

The curves and the Riemann surfaces have been found for the $A_{n}, B_{n}
,C_n, D_{n}$ and $G_2$ Lie groups \cite{SW,LA,DA,BR,ARG,ALI}.
Especially, in \cite{MAR}, the authors claimed a unified framework
for finding
the curves of $N=2$ SUSY YM theory for all gauge groups. However they do
not give any explicit form of
curves for the exceptional gauge groups and any physical checks on the curves.

In this article, we present a unified scheme for finding the hyperelliptic
curve of the $N=2$ SUSY YM theory with any Lie gauge group.
The paper is organised as follows:

In section two, we introduce our unified scheme and redrive
previously obtained results for classical Lie gauge groups and $G_2$.

In section three, we present the curves for exceptional Lie groups
$F_4,E_{6,7,8}$ and check their physical consistency.
\Section{Unified Scheme}

We will now construct a complex curve for any gauge group in the form
\be \l{CUR}
y^2=W^2(x)-\Lambda ^{2{\tilde h}} x^k
\ee
The power of $\Lambda$ which is twice the dual Coxeter number ${\tilde h}$,
is determined by $U(1)_R$ anomaly \cite{SEI}, and $k$ is determined by the
degree of $W(x)$, the classical curve of gauge group. The singularity
structure of classical theory must be encoded in $W(x)$, which is determined
by Weyl group and its discriminant vanishes on the walls of the Weyl chamber.
Proceeding in this manner, and requiring $W(x)$ should be Weyl invariant, we
propose
\be \l{W}
W(x)=\prod _i (x-\vec \lambda _i\cdot \vec a)^2,
\ee
where $\vec \lambda _i$ are non-zero weight charge vectors of the
fundamental representation of the gauge group with the least dimension.
The discriminant of (\ref{W})
is
\be \l{DIS}
\Delta _W=\prod _{i<j} \{(\vec \lambda _i-\vec \lambda _j)\cdot \vec a \}^2,
\ee
and we have moduli space singularities when $\Delta _W=0$.

Now, let us reproduce the results of classical gauge group
\cite{SW,LA,DA,BR,ARG} and $G_2$ \cite{ALI}.

\vspace*{5mm}
{\bf I. $A_n$ Series}
\vspace*{5mm}

We take the following Dynkin diagram for $SU(n+1)$

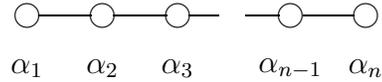
\begin{center}
\unitlength=1mm
\special{em:linewidth 0.4pt}
\linethickness{0.4pt}
\begin{picture}(80.00,13.00)
\put(10.00,10.00){\circle{3.00}}
\put(20.00,10.00){\circle{3.00}}
\put(30.00,10.00){\circle{3.00}}
\put(45.00,10.00){\circle{3.00}}
\put(55.00,10.00){\circle{3.00}}
\put(11.50,10.00){\line(2,0){7}}
\put(21.50,10.00){\line(2,0){7}}
\put(46.70,10.00){\line(2,0){6.80}}
\put(31.50,10.00){\line(2,0){4}}
\put(39.00,10.00){\line(2,0){4.5}}
\put(10.00,3.00){\makebox(0,0)[cc]{$\alpha_1$}}
\put(20.00,3.00){\makebox(0,0)[cc]{$\alpha_2$}}
\put(30.00,3.00){\makebox(0,0)[cc]{$\alpha_3$}}
\put(45.00,3.00){\makebox(0,0)[cc]{$\alpha_{n-1}$}}
\put(55.00,3.00){\makebox(0,0)[cc]{$\alpha_{n}$}}
\end{picture}
\end{center}

We choose the fundamental representation
$\Lambda _1$ which its dimension is $n+1$, so (\ref{W}) is
\be \l{WSUN}
W(x)=(x-a_1)(x-(a_2-a_1))\cdots(x-(a_n-a_{n-1}))(x+a_n),
\ee
and the quantum curve (\ref{CUR}) is
\be \l{CURSUN}
y^2=W^2(x)-\Lambda ^{2(n+1)},
\ee
which is the same curve as that obtained in \cite{LA}.

\vspace*{5mm}
{\bf II. $B_n$ Series}
\vspace*{5mm}

The Dynkin diagram of $SO(2n+1)$ is

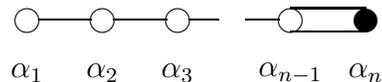
\begin{center}
\unitlength=1mm
\special{em:linewidth 0.4pt}
\linethickness{0.4pt}
\begin{picture}(80.00,13.00)
\put(10.00,10.00){\circle{3.00}}
\put(20.00,10.00){\circle{3.00}}
\put(30.00,10.00){\circle{3.00}}
\put(45.00,10.00){\circle{3.00}}
\put(55.00,10.00){\circle*{3.00}}
\put(11.50,10.00){\line(2,0){7}}
\put(21.50,10.00){\line(2,0){7}}
\put(45.00,11.60){\line(2,0){10}}
\put(45.00,8.45){\line(2,0){10}}
\put(31.50,10.00){\line(2,0){4}}
\put(39.00,10.00){\line(2,0){4.5}}
\put(10.00,3.00){\makebox(0,0)[cc]{$\alpha_1$}}
\put(20.00,3.00){\makebox(0,0)[cc]{$\alpha_2$}}
\put(30.00,3.00){\makebox(0,0)[cc]{$\alpha_3$}}
\put(45.00,3.00){\makebox(0,0)[cc]{$\alpha_{n-1}$}}
\put(55.00,3.00){\makebox(0,0)[cc]{$\alpha_{n}$}}
\end{picture}
\end{center}

We choose the fundamental representation
$\Lambda _1$ which its dimension is $2n+1$, so (\ref{W}) is
\be \l{WSOODD}
W(x)=(x^2-a_1^2)(x^2-(a_2-a_1)^2)\cdots(x^2-(a_n-a_{n-1})^2),
\ee
and the quantum curve (\ref{CUR}) is
\be \l{CURSOODD}
y^2=W^2(x)-\Lambda ^{(4n-2)}x^2.
\ee
It can easily be seen that by changing the basis of the root space
to orthogonal basis,
(\ref{CURSOODD}) is exactly the same curve as that obtained in \cite{DA}.

\vspace*{5mm}
{\bf III. $C_n$ Series}
\vspace*{5mm}

The Dynkin diagram for $SP(2n)$ group is
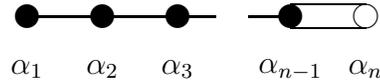
\begin{center}
\unitlength=1mm
\special{em:linewidth 0.4pt}
\linethickness{0.4pt}
\begin{picture}(80.00,13.00)
\put(10.00,10.00){\circle*{3.00}}
\put(20.00,10.00){\circle*{3.00}}
\put(30.00,10.00){\circle*{3.00}}
\put(45.00,10.00){\circle*{3.00}}
\put(55.00,10.00){\circle{3.00}}
\put(11.00,10.00){\line(2,0){11}}
\put(21.00,10.00){\line(2,0){11}}
\put(45.00,11.60){\line(2,0){10}}
\put(45.00,8.45){\line(2,0){10}}
\put(31.00,10.00){\line(2,0){4}}
\put(39.50,10.00){\line(2,0){4}}
\put(10.00,3.00){\makebox(0,0)[cc]{$\alpha_1$}}
\put(20.00,3.00){\makebox(0,0)[cc]{$\alpha_2$}}
\put(30.00,3.00){\makebox(0,0)[cc]{$\alpha_3$}}
\put(45.00,3.00){\makebox(0,0)[cc]{$\alpha_{n-1}$}}
\put(55.00,3.00){\makebox(0,0)[cc]{$\alpha_{n}$}}
\end{picture}
\end{center}

We choose the fundamental representation
$\Lambda _1$ which its dimension is $2n$, so (\ref{W}) is
\be \l{WSP2N}
W(x)=(x^2-a_1^2)(x^2-(a_2-a_1)^2)\cdots(x^2-(2a_n-a_{n-1})^2),
\ee
and the quantum curve (\ref{CUR}) is
\be \l{CURSP2N}
y^2=W^2(x)-\Lambda ^{2(n+1)}x^{2n-2},
\ee
which is the double cover of the curve obtained in \cite{ARG}.

\vspace*{5mm}
{\bf IV. $D_n$ Series}
\vspace*{5mm}

The Dynkin diagram of $SO(2n)$ is
\vspace*{10mm}
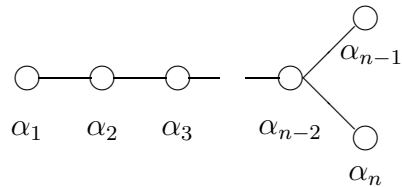
\begin{center}
\unitlength=1mm
\special{em:linewidth 0.4pt}
\linethickness{0.4pt}
\begin{picture}(80.00,13.00)
\put(10.00,10.00){\circle{3.00}}
\put(20.00,10.00){\circle{3.00}}
\put(30.00,10.00){\circle{3.00}}
\put(45.00,10.00){\circle{3.00}}
\put(55.00,18.00){\circle{3.00}}
\put(55.00,2.00){\circle{3.00}}
\put(11.50,10.00){\line(2,0){7}}
\put(21.50,10.00){\line(2,0){7}}
\put(46.70,10.00){\line(1,1){6.80}}
\put(46.70,10.00){\line(1,-1){6.80}}
\put(31.50,10.00){\line(2,0){4}}
\put(39.00,10.00){\line(2,0){4.5}}
\put(10.00,3.00){\makebox(0,0)[cc]{$\alpha_1$}}
\put(20.00,3.00){\makebox(0,0)[cc]{$\alpha_2$}}
\put(30.00,3.00){\makebox(0,0)[cc]{$\alpha_3$}}
\put(45.00,3.00){\makebox(0,0)[cc]{$\alpha_{n-2}$}}
\put(55.80,13.00){\makebox(0,0)[cc]{$\alpha_{n-1}$}}
\put(55.00,-3.00){\makebox(0,0)[cc]{$\alpha_{n}$}}
\end{picture}
\end{center}
\vspace*{10mm}

We choose the fundamental representation
$\Lambda _1$ which its dimension is $2n$, so (\ref{W}) is
\bea \l{WSO2N}
W(x)&=&(x^2-a_1^2)(x^2-(a_2-a_1)^2)\cdots
(x^2-(a_{n-2}-a_{n-3})^2) \cr
& &(x^2-(a_n+a_{n-1}-a_{n-2})) (x^2-(a_n-a_{n-1})^2),
\eea
and the quantum curve (\ref{CUR}) is
\be \l{CURSO2N}
y^2=W^2(x)-\Lambda ^{4(n-1)}x^{4},
\ee
which in a suitable orthogonal basis is exactly equivalent to the curve
given in \cite{BR}.

As we shall see, in contrast to the previous cases, there are additional
"singularities", we shall encounter the same situation for exceptional gauge
groups.
The discriminant of classical curve (\ref{WSO2N}) has the following
decomposition
\be \l{DELTASO2N}
\Delta _W=t^2\Delta,
\ee
in the notation of \cite{BR}, where $t^2=\prod _{i=1}^{2n} \vec
\lambda _i\cdot \vec a
$. The quantum discriminant has also a similar decomposition
\be \l{QDELTASO2N}
\Delta _\Lambda =t^4 \Delta ^{+} \Delta ^{-}.
\ee
The apparent "singularity" at $t=0$ is not a physical singularity
which means there is no massless dyon corresponding to it, i.e.
the monodromy
corresponding to encircling $t=0$, in the quantum moduli space, is trivial.
\cite{BR,ARG}.

\vspace*{5mm}

{\bf V. $G_2$ Case}
\vspace*{5mm}

The Dynkin diagram is

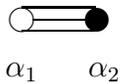
\begin{center}
\unitlength=1mm
\special{em:linewidth 0.4pt}
\linethickness{0.4pt}
\begin{picture}(80.00,13.00)
\put(30.00,10.00){\circle{3.00}}
\put(40.00,10.00){\circle*{3.00}}
\put(30.00,11.59){\line(2,0){9.78}}
\put(31.50,10.00){\line(2,0){8.50}}
\put(30.00,8.50){\line(2,0){9.70}}
\put(30.00,3.00){\makebox(0,0)[cc]{$\alpha_1$}}
\put(41.00,3.00){\makebox(0,0)[cc]{$\alpha_{2}$}}
\end{picture}
\end{center}

We choose the fundamental representation
$\Lambda _2$ which its dimension is $7$, so (\ref{W}) is
\be \l{WG2}
W(x)=(x^2-a_2^2)(x^2-(a_2-a_1)^2)(x^2-(a_1-2a_{2})^2),
\ee
and the quantum curve (\ref{CUR}) is
\be \l{CURG2}
y^2=W^2(x)-\Lambda ^8x^4,
\ee
which is the same as the curve given in \cite{ALI}.

\Section{Exceptional Lie Gauge Groups}

In this section, we apply our scheme to exceptional Lie gauge groups
$F_4,E_{6,7,8}$. In each cases, we present classical and quantum curves,
simple semiclassical
and exact monodromies and check that the simple semiclassical monodromies
can be obtained by the product of a pair of simple exact monodromies.
We also obtain quantum shift matrix for each case.

\vspace*{5mm}

{\bf I. $F_4$ Gauge Group}

\vspace*{5mm}
The group $F_4$ is of the rank $4$ and dimension $52$. The Dynkin
diagram of $F_4$ is

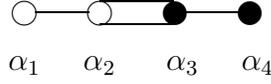
\begin{center}
\unitlength=1mm
\special{em:linewidth 0.4pt}
\linethickness{0.4pt}
\begin{picture}(80.00,13.00)
\put(20.00,10.00){\circle{3.00}}
\put(30.00,10.00){\circle{3.00}}
\put(40.00,10.00){\circle*{3.00}}
\put(50.00,10.00){\circle*{3.00}}
\put(21.50,10.00){\line(2,0){7}}
\put(30.00,11.60){\line(2,0){9.80}}
\put(30.00,8.50){\line(2,0){9.70}}
\put(41.50,10.00){\line(2,0){7}}
\put(20.00,3.00){\makebox(0,0)[cc]{$\alpha_1$}}
\put(30.00,3.00){\makebox(0,0)[cc]{$\alpha_2$}}
\put(41.00,3.00){\makebox(0,0)[cc]{$\alpha_{3}$}}
\put(51.00,3.00){\makebox(0,0)[cc]{$\alpha_{4}$}}
\end{picture}
\end{center}

The Weyl group of $F_4$ is the group of order $1152$ which is generated by

\bea \l{WEYLF4}
r_1&:&(a_1,a_2,a_3,a_4) \rightarrow (-a_1+a_2,a_2,a_3,a_4), \cr
r_2&:&(a_1,a_2,a_3,a_4) \rightarrow (a_1,a_1-a_2+2a_3,a_3,a_4), \cr
r_3&:&(a_1,a_2,a_3,a_4) \rightarrow (a_1,a_2,a_2-a_3+a_4,a_4), \cr
r_4&:&(a_1,a_2,a_3,a_4) \rightarrow (a_1,a_2,a_3,a_3-a_4).
\eea
We choose the fundamental representation
$\Lambda _4$ which its dimension is $26$, so (\ref{W}) is
\bea \l{WF4}
&W&(x)=
(x^2-a_4^2)(x^2-(a_4-a_3)^2)(x^2-(a_3-a_{2})^2)\cr
& &(x^2-(a_3-a_2+a_1)^2)(x^2-(a_3-a_1)^2)(x^2-(a_4-a_{3}+a_1)^2)  \\
& &(x^2-(a_4-a_1)^2)(x^2-(a_4-a_3+a_2-a_1)^2)(x^2-(a_4-a_{2}+a_1)^2) \cr
& &(x^2-(a_4+a_3-a_2)^2)(x^2-(a_4-2a_3+a_2)^2)(x^2-(2a_4-a_{3})^2)
, \nn
\eea
and the quantum curve (\ref{CUR}) is
\be \l{CURF4}
y^2=W^2(x)-\Lambda ^{18} x^{30}.
\ee
By our construction, the curve (\ref{CURF4}) is Weyl invariant and
hence could be
expressed in terms of Casimir invariants of the $F_4$,
that are $u_2,u_6,u_8,u_{12}$.
The classical discriminant of (\ref{WF4}) is factorized in the following form
\be \l{DELTAF4}
\Delta _W=f\prod _{\vec \alpha \in \delta ^{+} (F_4)}(\vec a \cdot \vec
\alpha ),
^2
\ee
where $f$ is a Weyl invariant function of $\vec a$. As the $D_n$ case,
there are unexpected "singularities"
where $f=0$.
We shall discuss them shortly.
The quantum discriminant is
\be \l{QDELTAF4}
\Delta _{\Lambda}=\prod _{i<j} (e_i^{+}-e_j^{+})^2
(e_i^{-}-e_j^{-})^2,
\ee
and can be factorized as follows
\be \l{QDELTAF4FAC}
\Delta _{\Lambda}=f^{+}f^{-}\Delta ^{+}\Delta ^{-}.
\ee
The hyperelleptic curve has $24$ cuts, and the homology cycles around them
are \be\ba{lll}    \l{VANF4}
\gamma _1=\a _4,&\gamma _2=\a _4-\a _3,&\gamma _3=\a _3-\a _2,\cr
\gamma _4=\a _3-\a _1,&\gamma _5=\a _4-\a _1,&\gamma _6=2\a_4-\a_3,
\cr
\gamma _7=\a_4+2\a_3-\a_2,&\gamma _8=\a _3-\a_2+\a _1,&\gamma _9=
\a _4+\a_3-\a_2,\cr
\gamma _{10}=\a_4-\a_2+\a _1,&\gamma _{11}=\a _4-\a _3+\a_1 ,&\gamma _{12}=
\a _4-\a _3+\a _2-\a_1,
\ea\ee
and $\gamma _{12+i}, i=1,\cdots,12$ are related to $\gamma_i$ by parity $x
\rightarrow -x$. The intersection requirements of the cycles $\beta _i$ with
$\a _i$, determine the $\beta _i$ to be
\be\l{VANF4D}
\gamma ^D_1=\beta _1+\beta _2+\beta _3+\beta _4,\quad
\gamma ^D_2=-\beta _1-\beta _2-\beta _3,\quad
\gamma ^D_3=-\beta _2,\quad
\gamma ^D_4=-\beta _1,
\ee
where $\gamma ^D_i$'s are the conjugate cycles to $\gamma _i$. As
$F_4$ is non-simply laced group, we must use the modified
Picard-Lefschetz formula for obtaining the exact monodromies \cite{DA,BR}.
In the quantum moduli space, $\gamma _i$'s do not vanish anywhere as far as
the semiclassical monodromies are concerned, unless $\Lambda =0$. For
$\Lambda \rightarrow 0$, the $\gamma _i$ cycles vanish, since $e_i^{+}
\rightarrow e_i^{-}$. Then it is straightforward to compute the monodromies
for the singularities at $\Lambda \rightarrow 0$, which we denote by
$B_i,\, i=1,\cdots ,12$ corresponding to each $\gamma _i$. By multiplying
these
matrices, one can obtain the quantum shift matrix,
$\prod _{i=1}^{12}B_i=T^{-3}$, where $T=\pmatrix{\bone & C\cr 0 &\bone }$,
and
\be \l{QSF4}
C=\pmatrix {2&-1&0&0 \cr
            -1&2&-2&0\cr
            0&-2&4&-2\cr
            0&0&-2&4},
\ee
in agreement with the shift obtained from the one loop corrected prepotential
under $\Lambda ^{18} \rightarrow e^{2\pi i\theta}\Lambda ^{18},\,\theta
\in [0,1]$. The simple semiclassical monodromies obtained from the one loop
corrected prepotential are
\be \l{M} M^{(r_i)}=\pmatrix{{(r_i^{-1})}^t&-C_i \cr 0&r_i}, \ee where
\be\ba{ll} \l{SEMCLASF4}
C_1=\pmatrix{-4&2&0&0\cr2&-1&0&0\cr0&0&0&0\cr0&0&0&0},&
C_2=\pmatrix{-1&2&-2&0\cr2&-4&4&0\cr-2&4&-4&0\cr0&0&0&0},\cr
&\cr
C_3=\pmatrix{0&0&0&0\cr0&-1&2&-1\cr0&2&-4&2\cr0&-1&2&-1},&
C_4=\pmatrix{0&0&0&0\cr0&0&0&0\cr0&0&-1&2\cr0&0&2&-4}.
\ea\ee
 To calculate exact monodromies, we look at the vanishing cycles of Riemann
surface where $\Delta ^{+}$ or $\Delta ^{-}=0$. We have chosen a slice in
moduli space
at fixed $u_4,u_6,u_{12}$ and varying $u_2$. Fixing a basis in this slice
and encircling the singularity, we obtain
the following simple monodromies
\be \ba{ll} \l{DYONF4}
M_1=M_{(1,0,0,0;0,0,0,0)},&M_2=M_{(1,0,0,0;-2,1,0,0)},\cr
M_3=M_{(0,1,0,0;0,0,0,0)},&M_4=M_{(0,1,0,0;-1,2,-2,0)},\cr
M_5=M_{(0,0,1,0;0,0,0,0)},&M_6=M_{(0,0,1,0;0,-1,2,-1)},\cr
M_7=M_{(0,0,0,1;0,0,0,0)},&M_8=M_{(0,0,0,1;0,0,-1,2)},
\ea \ee
where we use the usual notation
\be \l{EXACT}
M_{(\mm,\qq)}=\pmatrix{\bone +\qq \otimes \mm&\qq \otimes \qq \cr -\mm
\otimes \mm &\bone -
\mm \otimes \qq}.
\ee
The other monodromies can be obtained from the above simple monodromies
by the conjugation. The product of two strong monodromies in  (\ref{DYONF4})
generates weak monodromies (\ref{SEMCLASF4}) by
\be \l{STRONGWEAK}
M_1M_2=M^{(r_1)},
M_3M_4=M^{(r_2)},
M_5M_6=M^{(r_3)},
M_7M_8=M^{(r_4)}.
\ee
As we noted above, when in the classical discriminant (\ref{DELTAF4})
$f=0$ or
in the quantum discriminant (\ref{QDELTAF4}) $f^+=f^-=0$, there are
some "singularities". We argue here that at these "singularities"
no BPS saturated states become massless.
In fact, by encircling these "singularities", there is no change
in the logarithmic term of prepotential, hence there is not any quantum shift
for these classical monodromies. Therefore, it seems that in these
"singularities", no BPS saturated states become massless.
Moreover, by tracing the branch points of some of these
"singularities" for some special slices in moduli space, we have
checked that their paths do not intersect
any other path singularities and hence the monodromies associated
with these "singularities" are trivial.

\vspace*{5mm}
{\bf II. $E_6$ Gauge Group}
\vspace*{5mm}

The group $E_6$ is of rank $6$ and dimension $78$ and its Dynkin diagram is
\vspace*{5mm}
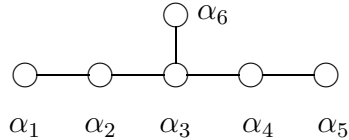
\begin{center}
\unitlength=1mm
\special{em:linewidth 0.4pt}
\linethickness{0.4pt}
\begin{picture}(80.00,13.00)
\put(10.00,10.00){\circle{3.00}}
\put(20.00,10.00){\circle{3.00}}
\put(30.00,10.00){\circle{3.00}}
\put(40.00,10.00){\circle{3.00}}
\put(50.00,10.00){\circle{3.00}}
\put(30.00,18.00){\circle{3.00}}
\put(11.50,10.00){\line(2,0){7}}
\put(21.50,10.00){\line(2,0){7}}
\put(31.50,10.00){\line(2,0){7}}
\put(41.50,10.00){\line(2,0){7}}
\put(30.00,11.50){\line(0,1){4.9}}
\put(10.00,3.00){\makebox(0,0)[cc]{$\alpha_1$}}
\put(20.00,3.00){\makebox(0,0)[cc]{$\alpha_2$}}
\put(30.00,3.00){\makebox(0,0)[cc]{$\alpha_3$}}
\put(41.00,3.00){\makebox(0,0)[cc]{$\alpha_{4}$}}
\put(51.00,3.00){\makebox(0,0)[cc]{$\alpha_{5}$}}
\put(35.00,18.00){\makebox(0,0)[cc]{$\alpha_{6}$}}
\end{picture}
\end{center}

The Weyl group of $E_6$ is of the order $51840$, which is generated by
\bea \l{WEYLE6}
r_1&:&(a_1,a_2,a_3,a_4,a_5,a_6) \rightarrow (-a_1+a_2,a_2,a_3,a_4,a_5,a_6),\cr
r_2&:&(a_1,a_2,a_3,a_4,a_5,a_6) \rightarrow (a_1,a_1-a_2+a_3,a_3,a_4,a_5,a_6),\cr
r_3&:&(a_1,a_2,a_3,a_4,a_5,a_6) \rightarrow (a_1,a_2,a_2-a_3+a_4+a_6,a_4,a_5
,a_6), \cr
r_4&:&(a_1,a_2,a_3,a_4,a_5,a_6) \rightarrow (a_1,a_2,a_3,a_3-a_4+a_5,a_5,a_6)
, \cr
r_5&:&(a_1,a_2,a_3,a_4,a_5,a_6) \rightarrow (a_1,a_2,a_3,a_4,a_4-a_5,a_6), \cr
r_6&:&(a_1,a_2,a_3,a_4,a_5,a_6) \rightarrow (a_1,a_2,a_3,a_4,a_5,a_3-a_6).
\eea
We choose the fundamental representation
$\Lambda _1$ which its dimension is $27$, so (\ref{W}) is
\bea \l{WE6}
&W&(x)=(x-a_1)(x-a_2+a_1)(x-a_5+a_2-a_1)(x-a_6-a_4+a_3) \cr
& &(x-a_6-a_5+a_4)(x+a_6-a_4)(x-a_6+a_5)(x+a_6-a_5+a_4-a_3) \cr
& &(x+a_6+a_5-a_3)(x-a_5+a_3-a_2)(x+a_5-a_4+a_3-a_2)(x-a_3+a_2) \cr
& &(x+a_4-a_2)(x+a_5-a_4+a_2-a_1)(x-a_5+a_1)(x+a_4-a_3+a_2-a_1) \cr
& &(x+a_5-a_4+a_1)(x-a_6+a_3-a_1)(x+a_4-a_3+a_1)(x+a_6-a_1) \cr
& &(x-a_6+a_3-a_2+a_1)(x+a_6-a_2+a_1)(x-a_6+a_2)(x+a_6-a_3+a_2) \cr
& &(x-a_4+a_3)(x-a_5+a_4)(x+a_5),
\eea
and the quantum curve (\ref{CUR}) is
\be \l{CURE6}
y^2=W^2(x)-\Lambda ^{24}x^{30}.
\ee
We can also express the above Weyl invariant curve in terms of the
Casimir invariants of
the group which are $u_2,u_5,u_6,u_8,u_9,u_{12}$.
The classical and quantum discriminants can
be factorized as the case of $F_4$ and hence we have a large number of
unphysical "singularities".
In this case, like the $F_4$ case, these "singularities" can not produce any
non-trivial monodromies, i.e. massless BPS saturated states.
The hyperelliptic curve (\ref{CURE6}) has $27$ cuts
that can easily be written from classical level surface \cite{LER}.
These $27$ cycles vanish when $\Lambda \rightarrow 0$, so one can
compute the monodromies for these singularities ($B_i,\,i=1,\cdots ,27$) and
obtain the quantum shift matrix
$\prod _{i=1} ^{27} B_i =T^{-6}$ where
$T=\pmatrix{\bone &C\cr 0&\bone}$ and $C$ is the Cartan matrix of $E_6$,
as we expected from one loop corrected prepotential.
The simple semiclassical monodromies obtained from the one loop
corrected prepotential are in the form (\ref{M}). Now the $C_i$'s are
\be \ba{ll} \l{SEMCLASSE6}
C_1=\pmatrix{-4&2&0&0&0&0\cr2&-1&0&0&0&0\cr0&0&0&0&0&0\cr
0&0&0&0&0&0\cr 0&0&0&0&0&0\cr 0&0&0&0&0&0},&
C_2=\pmatrix{-1&2&-1&0&0&0\cr2&-4&2&0&0&0\cr-1&2&-1&0&0&0\cr
0&0&0&0&0&0\cr 0&0&0&0&0&0\cr 0&0&0&0&0&0},\cr
&\cr
C_3=\pmatrix{0&0&0&0&0&0\cr0&-1&2&-1&0&-1\cr0&2&-4&2&0&2\cr
0&-1&2&-1&0&-1\cr 0&0&0&0&0&0\cr 0&-1&2&-1&0&-1},&
C_4=\pmatrix{0&0&0&0&0&0\cr0&0&0&0&0&0\cr0&0&-1&2&-1&0\cr
0&0&2&-4&2&0\cr 0&0&-1&2&-1&0\cr 0&0&0&0&0&0},\cr
&\cr
C_5=\pmatrix{0&0&0&0&0&0\cr0&0&0&0&0&0\cr0&0&0&0&0&0\cr
0&0&0&-1&2&0\cr 0&0&0&2&-4&0\cr 0&0&0&0&0&0},&
C_6=\pmatrix{0&0&0&0&0&0\cr0&0&0&0&0&0\cr0&0&-1&0&0&2\cr
0&0&0&0&0&0\cr 0&0&0&0&0&0\cr 0&0&2&0&0&-4}.
\ea \ee
Finally, to obtain the exact monodromies for $E_6$ group, we choose a slice
in moduli space at fixed $u_5,u_6,u_8,u_9,u_{12}$ and varying $u_2$, then
we get the simple exact monodromies
\be \ba{ll} \l{DYONE6}
M_1=M_{(1,0,0,0,0,0;0,0,0,0,0,0)},&M_2=M_{(1,0,0,0,0,0;2,-1,0,0,0,0)},\cr
M_3=M_{(0,1,0,0,0,0;0,0,0,0,0,0)},&M_4=M_{(0,1,0,0,0,0;-1,2,-1,0,0,0)},\cr
M_5=M_{(0,0,1,0,0,0;0,0,0,0,0,0)},&M_6=M_{(0,0,1,0,0,0;0,-1,2,-1,0,-1)},\cr
M_7=M_{(0,0,0,1,0,0;0,0,0,0,0,0)},&M_8=M_{(0,0,0,1,0,0;0,0,-1,2,-1,0)},\cr
M_9=M_{(0,0,0,0,1,0;0,0,0,0,0,0)},&M_{10}=M_{(0,0,0,0,1,0;0,0,0,-1,2,0)},\cr
M_{11}=M_{(0,0,0,0,0,1;0,0,0,0,0,0)},&M_{12}=M_{(0,0,0,0,0,1;0,0,-1,0,0,2)},
\ea \ee
and the other exact monodromies can be obtained by the conjugation of
the above basic
monodromies. We note that the product of any pairs of exact monodromies in
(\ref{DYONE6}) reproduce the semiclassical monodromies, i.e.
\be \ba {lll} \l{STRONGWEAKE6}
M_1M_2=M^{(r_1)},&
M_3M_4=M^{(r_2)},&
M_5M_6=M^{(r_3)}\cr
M_7M_8=M^{(r_4)},&
M_9M_{10}=M^{(r_5)},&
M_{11}M_{12}=M^{(r_6)}
\ea \ee

\vspace*{5mm}
{\bf III. $E_7$ Gauge Group}
\vspace*{5mm}

The group $E_7$ is of rank $7$ and dimension $133$ and its Dynkin diagram is
\vspace*{5mm}
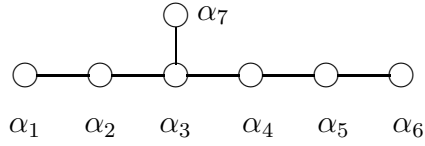
\begin{center}
\unitlength=1mm
\special{em:linewidth 0.4pt}
\linethickness{0.4pt}
\begin{picture}(80.00,13.00)
\put(10.00,10.00){\circle{3.00}}
\put(20.00,10.00){\circle{3.00}}
\put(30.00,10.00){\circle{3.00}}
\put(40.00,10.00){\circle{3.00}}
\put(50.00,10.00){\circle{3.00}}
\put(60.00,10.00){\circle{3.00}}
\put(30.00,18.00){\circle{3.00}}
\put(11.50,10.00){\line(2,0){7}}
\put(21.50,10.00){\line(2,0){7}}
\put(31.50,10.00){\line(2,0){7}}
\put(41.50,10.00){\line(2,0){7}}
\put(51.50,10.00){\line(2,0){7}}
\put(30.00,11.50){\line(0,1){4.9}}
\put(10.00,3.00){\makebox(0,0)[cc]{$\alpha_1$}}
\put(20.00,3.00){\makebox(0,0)[cc]{$\alpha_2$}}
\put(30.00,3.00){\makebox(0,0)[cc]{$\alpha_3$}}
\put(41.00,3.00){\makebox(0,0)[cc]{$\alpha_{4}$}}
\put(51.00,3.00){\makebox(0,0)[cc]{$\alpha_{5}$}}
\put(61.00,3.00){\makebox(0,0)[cc]{$\alpha_{6}$}}
\put(35.00,18.00){\makebox(0,0)[cc]{$\alpha_{7}$}}
\end{picture}
\end{center}

The Weyl group of $E_7$ is of the order $2903040$, that is generated by
\bea \l{WEYLE7}
r_1&:&(a_1,a_2,a_3,a_4,a_5,a_6,a_7) \rightarrow (-a_1+a_2,a_2,a_3,a_4,a_5,a_6
,a_7),\cr
r_2&:&(a_1,a_2,a_3,a_4,a_5,a_6,a_7) \rightarrow (a_1,a_1-a_2+a_3,a_3,a_4,a_5,a_6,
a_7),\cr
r_3&:&(a_1,a_2,a_3,a_4,a_5,a_6,a_7) \rightarrow (a_1,a_2,a_2-a_3+a_4+a_7,a_4,
a_5,a_6,a_7),\cr
r_4&:&(a_1,a_2,a_3,a_4,a_5,a_6,a_7) \rightarrow (a_1,a_2,a_3,a_3-a_4+a_5,a_5,
a_6,a_7),\cr
r_5&:&(a_1,a_2,a_3,a_4,a_5,a_6,a_7) \rightarrow (a_1,a_2,a_3,a_4,a_4-a_5+a_6
,a_6,a_7),\cr
r_6&:&(a_1,a_2,a_3,a_4,a_5,a_6,a_7) \rightarrow (a_1,a_2,a_3,a_4,a_5,a_5-a_6
,a_7),\cr
r_7&:&(a_1,a_2,a_3,a_4,a_5,a_6,a_7) \rightarrow (a_1,a_2,a_3,a_4,a_5,a_6,a_3
-a_7).\eea
We choose the fundamental representation
$\Lambda _6$ which its dimension is $56$, so (\ref{W}) is
\bea \l{WE7}
&W&(x)=
(x^2-a_6^2)(x^2-(a_5-a_6)^2)(x^2-(a_4-a_{5})^2)\cr
& &(x^2-(a_3-a_4)^2)(x^2-(a_2-a_3+a_7)^2)(x^2-(a_1-a_2+a_7)^2)\cr
& &(x^2-(a_2-a_7)^2)(x^2-(a_7-a_1)^2)(x^2-(a_1-a_2+a_3-a_7)^2)\cr
& &(x^2-(a_3-a_7-a_1)^2)(x^2-(-a_3+a_4+a_1)^2)(x^2-(-a_1+a_2-a_3+a_4)^2) \cr
& &(x^2-(a_5-a_4+a_1)^2)(x^2-(a_1-a_5+a_6)^2)(x^2-(-a_3+a_4-a_6+a_7)^2)  \cr
& &(x^2-(-a_1+a_2-a_4+a_5)^2)(x^2-(-a_1+a_2-a_5+a_6)^2)(x^2-(a_1-a_6)^2)  \cr
& &(x^2-(-a_2+a_3-a_4+a_5)^2)(x^2-(-a_1+a_2-a_6)^2)(x^2-(-a_3+a_5+a_7)^2)  \cr
& &(x^2-(-a_2+a_3-a_5+a_6)^2)(x^2-(-a_2+a_3-a_6)^2)(x^2-(a_2-a_4)^2)  \cr
& &(x^2-(a_4-a_6-a_7)^2)(x^2-(a_5-a_7)^2)(x^2-(-a_4+a_5-a_6+a_7)^2)  \cr
& &(x^2-(-a_3+a_4-a_5+a_6+a_7)^2),
\eea
and the quantum curve is
\be \l{CURE7}
y^2=W^2(x)-\Lambda^{36} x^{76},
\ee
that can be express in terms of Casimir invariants of $E_7$ which
are $u_2,u_6,u_8,u_{10},u_{12},$\newline $u_{14}$ and $u_{18}$.

Similar to $F_4$ and $E_6$ cases, the classical and quantum discreminant
can be factoraized and hence we have many "singularites" which are again
unphysical. The hyperelleptic curve (\ref{CURE7}) has $56$ cuts that
can be written from classical level surface.
These $56$ cycles vanish when $\Lambda \rightarrow 0$, so one can
compute the monodromies for these singularities ($B_i,\,i=1,\cdots ,28$) and
obtain the quantum shift matrix
$\prod _{i=1} ^{28} B_i =T^{-6}$ where
$T=\pmatrix{\bone &C\cr 0&\bone}$ and $C$ is the Cartan matrix of $E_7$, in
agreement with the one loop corrected prepotential.
The simple semiclassical monodromies obtained from the one loop
corrected prepotential are in the form (\ref{M}). Now $C_i$'s are
$$ \ba{ll}
C_1=\pmatrix{-4&2&0&0&0&0&0\cr2&-1&0&0&0&0&0\cr0&0&0&0&0&0&0\cr
0&0&0&0&0&0&0\cr 0&0&0&0&0&0&0\cr 0&0&0&0&0&0&0\cr 0&0&0&0&0&0&0},&
C_2=\pmatrix{-1&2&-1&0&0&0&0\cr2&-4&2&0&0&0&0\cr-1&2&-1&0&0&0&0\cr
0&0&0&0&0&0&0\cr 0&0&0&0&0&0&0\cr 0&0&0&0&0&0&0\cr 0&0&0&0&0&0&0}\cr
&\cr
C_3=\pmatrix{0&0&0&0&0&0&0\cr0&-1&2&-1&0&0&-1\cr0&2&-4&2&0&0&-2\cr
0&-1&2&-1&0&0&-1\cr 0&0&0&0&0&0&0\cr 0&0&0&0&0&0&0\cr 0&-1&2&-1&0&0&-1},&
C_4=\pmatrix{0&0&0&0&0&0&0\cr0&0&0&0&0&0&0\cr0&0&-1&2&-1&0&0\cr
0&0&2&-4&2&0&0\cr 0&0&-1&2&-1&0&0\cr 0&0&0&0&0&0&0\cr 0&0&0&0&0&0&0}
\cr &\cr
C_5=\pmatrix{0&0&0&0&0&0&0\cr0&0&0&0&0&0&0\cr0&0&0&0&0&0&0\cr
0&0&0&-1&2&-1&0\cr 0&0&0&2&-4&2&0\cr 0&0&0&-1&2&-1&0\cr 0&0&0&0&0&0&0},&
C_6=\pmatrix{0&0&0&0&0&0&0\cr0&0&0&0&0&0&0\cr0&0&0&0&0&0&0\cr
0&0&0&0&0&0&0\cr 0&0&0&0&-1&2&0\cr 0&0&0&0&2&-4&0\cr 0&0&0&0&0&0&0}
\ea $$
\be \ba{ll} \l{SEME7}
C_7=\pmatrix{0&0&0&0&0&0&0\cr0&0&0&0&0&0&0\cr0&0&-1&0&0&0&2\cr
0&0&0&0&0&0&0\cr 0&0&0&0&0&0&0\cr 0&0&0&0&0&0&0\cr 0&0&2&0&0&0&-4}.
\ea \ee
The simple exact monodromies can be obtained in the same way as previous
cases and are
\be \ba{ll}         \l{DYONE7}
M_1=M_{(1,0,0,0,0,0,0;0,0,0,0,0,0,0)},&M_2=M_{(1,0,0,0,0,0,0;2,-1,0,0,0,0,
0)},\cr
M_3=M_{(0,1,0,0,0,0,0;0,0,0,0,0,0,0)},&M_4=M_{(0,1,0,0,0,0,0;-1,2,-1,0,0,0,
0)},\cr
M_5=M_{(0,0,1,0,0,0,0;0,0,0,0,0,0,0)},&M_6=M_{(0,0,1,0,0,0,0;0,-1,2,-1,0,0,
-1)},\cr
M_7=M_{(0,0,0,1,0,0,0;0,0,0,0,0,0,0)},&M_8=M_{(0,0,0,1,0,0,0;0,0,-1,2,-1,0,
0)},\cr
M_9=M_{(0,0,0,0,1,0,0;0,0,0,0,0,0,0)},&M_{10}=M_{(0,0,0,0,1,0,0;0,0,0,-1,2,
-1,0)}, \cr
M_{11}=M_{(0,0,0,0,0,1,0;0,0,0,0,0,0,0)},&M_{12}=M_{(0,0,0,0,0,1,0;0,0,0,0,
-1,2,0)},\cr
M_{13}=M_{(0,0,0,0,0,0,1;0,0,0,0,0,0,0)},&M_{14}=M_{(0,0,0,0,0,0,1;0,0,-1,
0,0,0,2)},
\ea \ee
and the other exact monodromies can be obtained by the conjugation of
above basic
monodromies. We note that the product of any pairs of exact monodromies in
(\ref{DYONE7}) reproduces the semiclassical monodromies, i.e.
\be \ba {lll} \l{STRONGWEAKE7}
M_1M_2=M^{(r_1)},&
M_3M_4=M^{(r_2)},&M_5M_6=M^{(r_3)},\cr
M_7M_8=M^{(r_4)},&
M_9M_{10}=M^{(r_5)},&
M_{11}M_{12}=M^{(r_6)},\cr
M_{13}M_{14}=M^{(r_7)}.&
&

\ea \ee

\vspace*{5mm}
{\bf IV. $E_8$ Gauge Group}
\vspace*{5mm}

The group $E_8$ is of rank $8$ and dimension $248$ and its Dynkin diagram is
\vspace*{5mm}
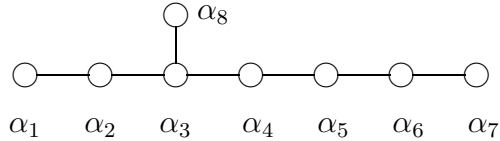
\begin{center}
\unitlength=1mm
\special{em:linewidth 0.4pt}
\linethickness{0.4pt}
\begin{picture}(80.00,13.00)
\put(10.00,10.00){\circle{3.00}}
\put(20.00,10.00){\circle{3.00}}
\put(30.00,10.00){\circle{3.00}}
\put(40.00,10.00){\circle{3.00}}
\put(50.00,10.00){\circle{3.00}}
\put(60.00,10.00){\circle{3.00}}
\put(70.00,10.00){\circle{3.00}}
\put(30.00,18.00){\circle{3.00}}
\put(11.50,10.00){\line(2,0){7}}
\put(21.50,10.00){\line(2,0){7}}
\put(31.50,10.00){\line(2,0){7}}
\put(41.50,10.00){\line(2,0){7}}
\put(51.50,10.00){\line(2,0){7}}
\put(61.50,10.00){\line(2,0){7}}
\put(30.00,11.50){\line(0,1){4.9}}
\put(10.00,3.00){\makebox(0,0)[cc]{$\alpha_1$}}
\put(20.00,3.00){\makebox(0,0)[cc]{$\alpha_2$}}
\put(30.00,3.00){\makebox(0,0)[cc]{$\alpha_3$}}
\put(41.00,3.00){\makebox(0,0)[cc]{$\alpha_{4}$}}
\put(51.00,3.00){\makebox(0,0)[cc]{$\alpha_{5}$}}
\put(61.00,3.00){\makebox(0,0)[cc]{$\alpha_{6}$}}
\put(71.00,3.00){\makebox(0,0)[cc]{$\alpha_{7}$}}
\put(35.00,18.00){\makebox(0,0)[cc]{$\alpha_{8}$}}
\end{picture}
\end{center}

The Weyl group of $E_8$ is of the order $696729600$, that is generated by
eight simple reflections corresponding to simple roots. Because of the
complexity
of the results, here we only list some of them.
We choose the fundamental representation
$\Lambda _7$ which is $248$ dimensional, so (\ref{W}) is
\be \l{WE8}
W(x)=\prod _{i=1}^{120}(x^2-b_i^2)
\ee
where $b_i$'s are given in Appendix.
The Weyl invariant quantum curve is in the following form
\be \l{CURE8}
y^2=W^2(x)-\Lambda^{60} x^{420},
\ee
that can be express in terms of the Casimir invariants of $E_8$
which are
$u_2,u_8,u_{12},u_{14},$\newline $u_{18},u_{20},u_{24},u_{30}$.

Similar to the previous exceptional gauge groups, the classical and quantum
discriminants can be factoraized and we have many "singularities" which
seem
unphysical. The hyperelliptic curve (\ref{CURE8}) has $240$ cuts, that half
of them are related to the other half by the parity.
These $120$ cycles vanish when $\Lambda \rightarrow 0$, so one can
compute the monodromies for these singularities ($B_i,\,i=1,\cdots ,120$) and
then the quantum shift matrix becomes in the form
$T=\pmatrix{\bone &C\cr 0&\bone}$ that $C$ is the Cartan matrix of $E_8$.
The simple semiclassical monodromies obtained from the one loop
corrected prepotential are in the form (\ref{M}).
The simple exact monodromies can be obtained in the same way as the
previous cases
and the results are
\be \ba{ll} \l{DYONE8}
M_1=M_{(1,0,0,0,0,0,0,0;0,0,0,0,0,0,0,0)},&M_2=M_{(1,0,0,0,0,0,0,0;2,-1,0,0,
0,0,0,0)},\cr
M_3=M_{(0,1,0,0,0,0,0,0;0,0,0,0,0,0,0,0)},&M_4=M_{(0,1,0,0,0,0,0,0;-1,2,-1,
0,0,0,0,0)},\cr
M_5=M_{(0,0,1,0,0,0,0,0;0,0,0,0,0,0,0,0)},&M_6=M_{(0,0,1,0,0,0,0,0;0,-1,2,
-1,0,0,0,-1)},\cr
M_7=M_{(0,0,0,1,0,0,0,0;0,0,0,0,0,0,0,0)},&M_8=M_{(0,0,0,1,0,0,0,0;0,0,-1,
2,-1,0,0,0)},\cr
M_9=M_{(0,0,0,0,1,0,0,0;0,0,0,0,0,0,0,0)},&M_{10}=M_{(0,0,0,0,1,0,0,0;0,0,0,
-1,2,-1,0,0)} ,\cr
M_{11}=M_{(0,0,0,0,0,1,0,0;0,0,0,0,0,0,0,0)},&M_{12}=M_{(0,0,0,0,0,1,0,0;0,
0,0,0,-1,2,-1,0)},\cr
M_{13}=M_{(0,0,0,0,0,0,1,0;0,0,0,0,0,0,0,0)},&M_{14}=M_{(0,0,0,0,0,0,1,0;0,
0,0,0,0,-1,2,0)},\cr
M_{15}=M_{(0,0,0,0,0,0,0,1;0,0,0,0,0,0,0,0)},&M_{16}=M_{(0,0,0,0,0,0,0,1;0,
0,-1,0,0,0,0,2)},
\ea \ee
and the other exact monodromies can be obtained by the conjugation of the
above basic
monodromies. One can see that by multiplication of any pairs of exact
monodromies in
(\ref{DYONE8}), the semiclassical monodromies are reproduced.

\Section{Conclusions}

In this article, we have presented the hyperelliptic curves for any Lie
gauge group.
We have found that they have the form $y^2=W^2(x)-\Lambda ^{2\hat h}x^k$
where $W(x)$
is a polynomial of order $n$ ( $n$ is the dimension of the
fundamental
representation with least dimension minus the number of zero weights)
, $\hat h$ is the dual Coxeter number of the Lie gauge group and $k<2n$.
It has been found in \cite{DA}
that the one form $\lambda$ for these hyperelliptic curves is
$\lambda =({k \o 2}W-xW'){{dx} \o y}$, and we have been found that the
simple exact
monodromies can be obtained only from the Cartan matrix ($C=\{C_{ij}\}$).
In this basis, we have
$r=rank\,(G)$ massless monopoles \cite{LA}. The simple exact monodromies
are in
the form $M_{(e_i,0)}$ and $M_{(e_i,\sum _{j=1}^rC_{ij}e_j)},\,i=1,\cdots ,r$
, for every simple root of $G$, where $e_i$ is the unit vector in the $i$th
direction of Cartan subspace. The simple semiclassical monodromies
can also be obtained from the rows of the Cartan matrix. The other
monodromies
can be obtained by the conjugation.

It is interesting to note that by deleting some vertex in Dynkin diagram
of the gauge group $G$, one can find the hyperelliptic curve of the related
subgroup of $G$ from the curve of $G$\cite{DA}. For example, by deleting the
vertex
$\alpha _1$ or $\alpha _4$ from the Dynkin diagram of $F_4$, one can find
the curve of $SP(6)$ or $SO(7)$ respectively from the hyperelliptic
curve (\ref{CURF4}) of $F_4$. Similar results can be obtained
from the other exceptional curves.

\vspace*{5mm}
{\large ACKNOWLEDGEMENTS}
\vspace*{5mm}

We would like to thank Farhad Ardalan, Shahrokh Parvizi
for helpful discussions. M. A. also wishes to thank Wolfgang Lerche for
his useful comments and discussions.

\vspace*{5mm}
{\large APPENDIX}
\vspace*{5mm}

Here we list the $b_i$'s factor appearing in the classical curve (\ref{WE8})
$$ \ba{ll}
b_1=\7,&b_2=\7-\6,\\b_3=\6-\5,&
b_4=\5-\4,\\b_5=\4-\3,&b_6=\8-\3+\2,\\
b_7=\8-\2+\1,&b_8=\8-\2,\\b_9=\8-\1,&
b_{10}=\8-\3+\2-\1,\\b_{11}=\8-\3+\1,&b_{12}=\4-\3+\1,\\
b_{13}=\4-\3+\2-\1,&b_{14}=\5-\4+\1,\\
b_{15}=\4-\2,&
b_{16}=\5-\4+\2-\1,\\b_{17}=\6-\5+\1,&b_{18}=\5-\4+\3-\2,\\
b_{19}=\6-\5+\2-\1,&b_{20}=\7-\6+\1,\\b_{21}=\8+\5-\3,&
b_{22}=\6-\5+\3-\2,\\b_{23}=\7-\6+\2-\1,&b_{24}=\7-\1,\\
b_{25}=\8-\5,&b_{26}=\8+\6-\5+\4-\3,\\b_{27}=\7-\6+\3-\2,&
b_{28}=\7-\2+\1,\\b_{29}=\8-\6+\5-\4,&b_{30}=\8+\6-\4,\\
b_{31}=\8+\7-\6+\4-\3,&b_{32}=\7-\3+\2,\\b_{33}=\8-\6+\4-\3,&
b_{34}=\8+\7-\6+\5-\4,\\b_{35}=\8-\7+\6-\4,&b_{36}=\8-\7+\4-\3,\\
b_{37}=\6-\3+\2,&b_{38}=\8-\7+\6-\5+\4-\3,
\\b_{39}=\8+\7-\5,&
b_{40}=\8-\7+\5-\4,\\
b_{41}=\8+\7-\4,&b_{42}=\6-\2+\1,\\
b_{43}=\7-\6+\5-\3+\2,&b_{44}=\8+\7-\5+\4-\3,\\b_{45}=\8-\7+\5-\3,&
b_{46}=\7-\6+\5-\2+\1,\\b_{47}=\6-\1,&b_{48}=\8-\7+\6-\5,\\
b_{49}=\7-\5+\3-\2,&b_{50}=\7-\5+\4-\3+\2,\\b_{51}=\8-\6,&
b_{52}=\8+\7-\6+\5-\3,\\b_{53}=\7-\6+\5-\1,&b_{54}=\7-\5+\2-\1,\\
b_{55}=\7-\5+\4-\2+\1,&b_{56}=\7-\4+\2,\\b_{57}=\8+\6-\3,&
b_{58}=\7-\6+\5-\4+\3-\2,\\b_{59}=\7-\5+\1,&b_{60}=\7-\5+\4-\1,\\
b_{61}=\7-\4+\3-\2+\1,&b_{62}=\7-\6+\5-\4+\2-\1,
\ea $$
$$ \ba{ll}
b_{63}=\7-\6+\4-\2,&
b_{64}=\6-\4+\3-\2,\\b_{65}=\8+\7-\3+\1,&b_{66}=\7-\4+\3-\1,\\
b_{67}=\7-\6+\5-\4+\1,&b_{68}=\7-\6+\4-\3+\2-\1,\\b_{69}=\6-\5+\4-\2,&
b_{70}=\8+\7-\3+\2-\1,\\b_{71}=\8-\7-\1,&b_{72}=\6-\4+\2-\1,\\
b_{73}=\7-\6+\4-\3+\1,&b_{74}=\8-\7+\6-\3+\1,\\b_{75}=\5-\2,&
b_{76}=\6-\5+\4-\3+\2-\1,\\b_{77}=\6-\4+\1,&b_{78}=\8-\7-\2+\1,\\
b_{79}=\8-\7+\6-\3+\2-\1,&b_{80}=\8+\7-\6-\1,\\
b_{81}=\8+\7-\2,&
b_{82}=\6-\5+\4-\3+\1,\\b_{83}=\8-\6+\5-\3+\1,&b_{84}=\5-\3+\2-\1,\\
b_{85}=\8-\7-\3+\2,&b_{86}=\8+\7-\6-\2+\1,\\b_{87}=\8-\7+\6-\2,&
b_{88}=\8-\6+\5-\3+\2-\1,\\b_{89}=\8-\5+\4-\3+\1,&b_{90}=\8+\6-\5-\1,\\
b_{91}=\8+\7-\6-\3+\2,&b_{92}=\7+\4-\3,\\b_{93}=\5-\3+\1,&
b_{94}=\8+\6-\5-\2+\1,\\b_{95}=\8-\4+\1,&b_{96}=\8+\5-\4-\1,\\
b_{97}=\8-\6+\5-\2,&b_{98}=\8-\5+\4-\3+\2-\1,\\b_{99}=\7+\5-\4,&
b_{100}=\8+\6-\5-\3+\2,\\b_{101}=\8+\4-\3-\1,&b_{102}=\7-\6-\4+\3,\\
b_{103}=\8+\5-\4-\2+\1,&b_{104}=\8-\4+\2-\1,\\b_{105}=\8-\5+\4-\2,&
b_{106}=\7+\6-\5,\\b_{107}=\7-\6-\5+\4,&b_{108}=\3-\2-\1,\\
b_{109}=\8+\5-\4-\3+\2,&b_{110}=\8+\4-\3-\2+\1,\\b_{111}=\6-\5-\4+\3,&
b_{112}=\8-\4+\3-\2,\\b_{113}=2\7-\6,&b_{114}=\7-2\6+\5,\\
b_{115}=\2-2\1,&b_{116}=\6-2\5+\4,\\b_{117}=\3-2\2+\1,&
b_{118}=\5-2\4+\3,\\b_{119}=\8+\4-2\3+\2,&b_{120}=2\8-\3.
\ea $$


\begin{thebibliography}{99}
\bibitem{SW}
N. Seiberg and E. Witten, Nucl. Phys. B426 (1994) 19.

N. Seiberg and E. Witten, Nucl. Phys. B431 (1994) 484.

\bibitem{LA}
A. Klemm, W. Lerche, S. Yankielowicz and S. Theisen,
Phys. Lett. B344 (1995) 169.

A. Klemm, W. Lerche, and S. Theisen,
Int. J. Mod. Phys. A11 (1996) 1929.

P. Argyres, A. Faraggi, Phys. Rev. Lett. 73 (1995) 3931.

\bibitem{DA}
U. H. Danielsson, B. Sundborg, Phys. Lett. B358 (1995) 273.

\bibitem{BR}
A. Brandhuber, K. Landsteiner, Phys. Lett. B358 (1995) 73.

\bibitem{ARG}
P. C. Argyres, A. D. Shapere, RU-95-61, hep-th/9509175.

\bibitem{ALI}
M. Alishahiha, F. Ardalan, F. Mansouri, IPM-95-117, hep-th/9512005, to
be appear in Phys. Lett. B.

U. H. Danielsson, B. Sundborg, hep-th/9511180.

\bibitem{HAN}
A. Hanany, Y. Oz, Preprint TAUP-2248-95, hep-th/9505075.

A. Hanany, hep-th/9509176.

\bibitem{ALV}
L. Alvarez-Gaume, Introduction to $S$-duality, a Pedagogical Review of
Seiberg and Witten's Work, CERN, CH-1211 Geneva 23.

\bibitem{MAR}
E. Martinec, N. Warner, Preprint EFI-95-61, USC-95/025, hep-th/9509161.

\bibitem{SEI}
N. Seiberg, Phys. Lett. B209 (1988) 75.

\bibitem{LER}
W. Lerche, Preprint CERN-TH/95-183, hep-th/9507011.

\end{thebibliography}
\end{document}